# The characteristics of the superconducting and magnetic phases in the polycrystalline samples of ruthenocuprates of nominal compositions $RuSr_2GdCu_2O_8$ , $Ru_{0.98}Sr_2GdCu_2O_8$ and $Ru_{0.5}Sr_2GdCu_{2.5}O_{8-\delta}$

Piotr W Klamut [1] , Tomasz Plackowski, Marcin Matusiak

Institute of Low Temperature and Structure Research, Polish Academy of Sciences,

P Nr 1410, 50-950 Wrocław 2, Poland

## Abstract

The temperature dependencies of the resistivity for the superconducting ruthenocuprates of nominal compositions $RuSr_2GdCu_2O_8$, $Ru_{0.98}Sr_2GdCu_2O_8$ and $Ru_{0.5}Sr_2GdCu_{2.5}O_{8-d}$ were examined for the magnetic field dependent characteristics of the superconducting transitions. The effect of the insignificant diminishing of the Ru/Cu ratio in parent $RuSr_2GdCu_2O_8$ was confirmed as relevant for the stabilisation of the superconducting phase. Noted differences in the compared characteristics are interpreted for possible inhomogeneous nucleation of the superconducting phase in the parent ruthenocuprate. The phase anisotropy in $RuSr_2GdCu_2O_8$ and $Ru_{0.98}Sr_2GdCu_2O_8$ , in presence of the compounds Ru magnetism, appears to be a cause of a significant softening of the $H_{c2}(T)$ phase line. An anomalous lowering of the magnetoresistivity was observed in the approx. 10 K range above the onset of the superconducting transition, which may suggest the presence of enhanced superconducting fluctuations in the samples.

The positive magnetic field shift of the temperatures, which limit the magnetoresistivity and the specific heat signatures of the magnetic ordered state of the Ru sublattice, suggests probing the influence of the ferromagnetic Ru interactions in an effective metallic-like conduction channel present in the samples.

Superconducting characteristics of the $Ru_{0.5}Sr_2GdCu_{2.5}O_{8-d}$ reveal a significant contribution of the Gd paramagnetic signal at low temperatures, interpreted for the presence of a significant anisotropy of the superconducting phase. It is concluded that the Ru-Cu substituted phases of ruthenocuprates may present an opportunity to investigate the effectively anisotropic superconducting phase despite its comparatively high $T_c$ in the compounds related to the 123-type cuprate superconductor.

[1]corresponding author, e-mail: P.Klamut@int.pan.wroc.pl

**Introduction**

Interest in the properties of ruthenocuprates spanned several years of research being, in part, stimulated by an apparent coexistence of the compounds' magnetism with superconductivity. The magnetic and superconducting characteristics of these materials were, however, also found indicative of the occurrence of phase separation phenomena [1-3]. The particular microscopic character of such a separation, found to be most likely of structural origin, and the details of mutual influence of the magnetic and superconducting phases remain in the current research focus. Note, that the $RuSr_2GdCu_2O_8$ was originally reported non-superconducting [4] and the superconductivity may be induced in the compound on the particular synthesis route or in effect of its post synthesis annealing. Interestingly enough, in a course of previously reported investigations, such processes were not identified to associate any particular alteration of structural parameters or the overall stoichiometry of the compound, which could then be held for controlling the charge doping in the segments of the layered crystal structure, that in resemblance to many other high temperature superconducting (HTS) cuprates. Instead, the nanometre range alterations found in the crystal structure of the superconducting $RuSr_2GdCu_2O_8$ [5,6] suggest that such structural features may effectively influence, if not delineate, triggering of the superconducting properties in this compound. The superconductivity was also reported in the single crystals of $RuSr_2GdCu_2O_8$ with the *I-V* characteristics indicative of the Josephson junctions governed transport along the *c* axis of the crystal, which points to the atomic scale layered, however to an extend electronically separated, structure [6]. The mutual influence of the magnetism and superconductivity in the layered hybrids forms a broad and demanding subject of research to which the ruthenocuprates might indeed contribute with a sense of importance given to their nano-scaled features. Manifold properties, which were investigated for several families of the ruthenocuprate compounds were addressed in several review texts , *i.a.* in Refs. [1, 7-11].

It is of common intuition to the cuprates to consider relevance of the substitution driven hole doping control over the $T_c$. For $RuSr_2GdCu_2O_8$ ,however, so different values of the reported $T_c$ with superconductivity onset ranging in between 0 and approximately 50 K, suggest that subtle alterations of the overall stoichiometry may rather associate highly non-uniform structural alternation or affect involved magnetism in the compound. Increase of the superconducting $T_c$ was also observed in $Ru_{0.9}Sr_2GdCu_{2.1}O_8$ *vs.* nominally stoichiometric sample [12] and the maximum value of $T_c$ was reported at 72 K upon further Cu -> Ru substitution in the compound synthesised at high pressure [13]. The $T_c$ for $Ru_{1-x}Sr_2GdCu_{2+x}O_{8-d}$ , x>0, becomes dependent on the oxygen content [14], which resembles the well known oxygen content driven charge doping in the structurally related 123-type superconducting cuprates. The process of formation of the $RuSr_2GdCu_2O_8$ was recently studied in the experiments in which the amount of the Ru ions available for synthesis was varied by differing the chemical vapour transport of the Ru-oxides into the reaction zone [15, 16]. The resulting materials revealed slight changes in the Ru/Cu ratio and for such a modified phase it was noted that the increase of $T_c$ associates a slight decrease in the Ru content. The finite rate of diffusion of the Ru ions into the bulk, and simultaneous evaporative loss of the Ru atoms, were identified the two processes leading to different values of $T_c$ in the surface layer versus the core of the sample. Then, the material with homogeneous $T_c$ was reported by adjusting the efficiency of the chemical vapour transport [16]. In the associated investigation [17], for having established mixed valence, i.e. 4+ and 5+ valence states, of the Ru ions, it was proposed that a phase separation to ferromagnetic and antiferromagnetic domains (see also [1]) may be driven by competence between the $Ru^{4+}$-$Ru^{5+}$ double exchange and the $Ru^{5+}$- $Ru^{5+}$ super exchange interactions.

In the paper we comment on the characteristics of the superconducting transitions, which were measured for the polycrystalline samples with nominal compositions $RuSr_2GdCu_2O_8$, $Ru_{0.98}Sr_2GdCu_2O_8$ and $Ru_{0.5}Sr_2GdCu_{2.5}O_{8-d}$ . The superconducting $T_c$ ranges there from 0 to approximately 70 K depending on the Ru/Cu ratio in the samples, but also on the oxygen content as it was evidenced only for the third Cu-Ru substituted phase. We also report on the magnetic field dependencies of the resistive anomalies, which accompany the magnetic ordering temperatures $T_m^{Ru}$ in the parent phase. The nominal compositions $RuSr_2GdCu_2O_8$ and $Ru_{0.98}Sr_2GdCu_2O_8$ were chosen for investigating of the role of structural modifications induced by the nominal sub-stoichiometry of Ru in the 1212-type phase.

Several properties of these samples were discussed in the preceding articles [11, 18] where we extensively comment on the compounds synthesis, report the magnetic field dependencies of the isothermal magnetocaloric coefficient, and comment on the magnetic field dependent specific heat and magnetoresistivity of the normal state. The higher superconducting $T_c$ (throughout this text the $T_c$ references to an intra-grain superconducting transition) of the sample B ($Ru_{0.98}Sr_2GdCu_2O_8$, $T_c^{on} \approx$ 50 K) comparing to A ($T_c^{on} \approx$ 40 K $RuSr_2GdCu_2O_8$ ) was ascribed to the effect of deficiency of Ru written into the nominal formula. In [18] we also note that such induced structural modifications were programmed for gaining the comparative control over subtle alterations of the crystal structure, and they would not necessarily reflect the exact resultant stoichiometry of the sample. It should be noted that the synthesis of $RuSr_2GdCu_2O_8$ and $Ru_{0.98}Sr_2GdCu_2O_8$ samples was performed at high temperature close to the melting temperature for this phase, then at the conditions at which subtle modification of stoichiometry may occur due to enhanced sublimation of the Ru rich species. Overall it suggests that the achieved modification may only be enhanced, not induced, if only allowed by the underlying chemistry of the phase formation or for finite kinetics of the involved processes. Having no effective means for the microscopic control of such defects, we stress that the synthesis of samples A and B was performed at analogous conditions and simultaneously. Standard powder X-ray diffraction scans for these samples revealed, within accuracy of the measurement, neither a meaningful difference in the lattice parameters nor the presence of impurities (such scans are provided in the related Ref. [18]).

**Experimental**

The polycrystalline samples were prepared in the solid state reaction. The powders of $RuO_2$ (99.9% purity, preheated in 600°C), $Gd_2O_3$ (99.9% purity), $SrCO_3$ (99.9% purity) and CuO (99.9% purity) were calcined in air at 920°C for 53h with intermediate grinding. For $RuSr_2GdCu_2O_8$ and $Ru_{0.98}Sr_2GdCu_2O_8$ the product was sintered in the controlled flow of Ar at 1015 °C for 20h , which resulted in the fine mixture of the $Ru_{1-x}Sr_2GdO_6$ , $Cu_2O$ and eventual traces of other metal oxide phases. This step of the reaction prevents the formation of the secondary phase of strontium ruthenate perovskite [7], which when formed could be dissolved into the final reaction product only at high temperatures and with slow kinetics. The final compositions of the 1212-type phase were then sintered at 1060 °C in the flow of oxygen with intermediate cooling, grinding and pelletizing, with the product cooled at a rate of 1°C/min in the controlled flow of oxygen, then removed from the furnace at 600°C. Progress of the phase formation was monitored with the powder XRD scans. The samples $RuSr_2GdCu_2O_8$ (sample A) and $Ru_{0.98}Sr_2GdCu_2O_8$ (sample B) were synthesised simultaneously with the care being taken to assure exactly the same conditions for the reaction and the same handling of the material during whole process. Synthesis of these materials was extensively commented in [18], where more synthesis details are provided. The sample C of $Ru_{0.98}Sr_2GdCu_2O_8$ was prepared at the comparable conditions, although separately.

Synthesis of $Ru_{0.5}Sr_2GdCu_{2.5}O_{8-d}$ started from the stoichiometric mixture of the same oxides and strontium carbonate. After calcination in air at 920°C the product was reground, pressed, and annealed at 970 °C in the flowing oxygen. This annealing resulted in a multiphase material, then sintered for a period of 10 h at approximately 1115 °C at the high isostatic pressure of 3kbar in the oxidizing atmosphere of 21% $O_2$ (balance $N_2$) and slowly cooled in the closed gas volume. This sample was named sample D. Its part was then annealed in the flowing Ar gas at 650°C - sample E. Part of the sample E was finally re-annealed in the flow of oxygen at 600°C for 50 h, with slow cooling at a rate of 2°C/min - sample F.

The temperature dependencies of the *dc* resistivity were measured by the four contact method with 1mA *dc* current passed through the bar shaped (approx. 5 mm x 1-2 $mm^2$) samples. For such measurements the temperature was swept at 1°C/min, resulting in the temperature error for the single resistivity measurement to be less than 0.1 K. The data was collected on heating and cooling. For these measurements the magnetic field was set in the range between 0 and 130 kOe with use of the commercial Oxford Instruments magnet in the Teslatron setup. Temperature dependencies of the specific heat were measured with the relaxation method using the commercial micro-calorimeter probe in the PPMS platform system, by Quantum Design Inc. The temperature dependencies of the ac susceptibility (not shown here, referenced to [18]) were measured in the PPMS platform with the ac susceptibility probe. The temperature and field dependencies of the dc magnetisation were measured with the MPMS SQUID magnetometer, also by Quantum Design Inc. The electron microscopy analysis was performed with the EDAX equipped Hitachi S-4700-II Scanning Electron Microscope.

**Results and discussion**

Figure 1 presents the magnetic field dependencies of the characteristic temperatures which were chosen for describing the resistive superconducting transitions in the $Ru_{0.98}Sr_2GdCu_2O_8$ (sample A) and the $RuSr_2GdCu_2O_8$ (sample B). The temperature dependencies of the resistivity were measured for magnetic field values ranging from 0 to 130 kOe and such bare dependencies were presented in Ref. 11 (there see Fig. 3). The range of the onset of the transition was characterised with two temperatures, $T_c^{on}$ and $T_c^{on\ fl}$. The $T_c^{on}$ mark the onsets of significant increase of the derivative of resistivity and may be considered for the transition onset temperature. Higher in value $T_c^{on\ fl}$ corresponds to the lower limiting temperature for fitting the derivative of resistivity with one functional dependence above the transition –which was found to be linear for the samples A and B and constant for the samples D and F. The corresponding temperatures for the samples D and F are presented in Fig. 2. The $T_c^{on\ fl}$ may then correspond to the onset of superconducting fluctuations or reflect spread in the local values of $T_c$ , which may arise due to the phase inhomogeneity. It may also reflect the effect of magnetic field on the sample magnetism, for that influencing the superconducting response. The $T_c^{R=0}$ marks the temperature at which the resistivity attains zero value. At $T^{90\%Rn}$ the resistivity falls by 10% from its value at $T_c^{on}$, which may be used for conservative estimation of the phase line for the purpose of comparing $dH_{c2}(T)/dT$ values for different samples. For investigating these polycrystalline materials we constrain to comment only qualitative changes of such estimated phase lines, rather than arguing about the exact magnetic field dependences of the upper critical field.

The resistivity temperature dependencies, when compared with those typically reported for other polycrystalline superconducting cuprates, show that the resistive transitions in ruthenocuprates present rather broad, as seen in Figures 1 and 2 by assessing the temperature interval between $T_c^{on}$ and $T_c^{R=0}$ . This feature seems to involve comparatively weak links which are formed in granular structure of the ruthenocuprate sample (note the slope of the $T_c^{R=0}$ *(H)* in limit of the small magnetic field). The transition onset temperatures, as determined in the magnetic measurements, also remain very sensitive to the magnetic field. Then, comparatively the most accurate determination of the magnetic onset temperature of the superconducting transition comes by probing the incipient magnetic shielding in the dynamic (ac) susceptibility when it is measured at a limit of small amplitudes of the sensing field. For ruthenocuprates, such a determined onset is however still lower by several Kelvin from that determined in the resistivity measurement and it remains significantly shifted by changing the amplitude of sensing field in its limiting resolution of tenths of Oe – note that for samples A and B changing the ac field amplitude from 0.1 Oe to 1 Oe leads to lowering the $T_c^{on}$ as determined in the ac susceptibility by 2 K and 5 K respectively (our measured ac susceptibility-temperature dependences for the samples A and B are not shown here, they were presented in Ref. 11 , there Fig. 4). The described features may in principle be supported by spatial inhomogeneity of the superconducting phase, then with a few Kelvin different values of $T_c^{on}$, which may become further effectively enhanced for spatial inhomogoneity in the compound magnetism. Part of the literature proposed the picture of effective separation to the mesoscopic in size ferromagnetic/non-superconducting and the superconducting regions. Results of several experiments [5,6] suggest the consideration of the consequences of superconducting/magnetic layered hybrid structure. Note that for such, an inhomogeneity of the internal magnetic field, particularly of its component vertical to the superconducting layers, may well turn a driving force for inhomogeneous nucleation of the superconducting order parameter. The vertical component of the second critical field is in such structures inversely proportional to product of the superconducting coherence length and the thickness of the layer, and becomes significantly lowered in the limiting case of the thin film and temperatures in proximity of $T_c^{on}$. Previously reported analyses of the temperature dependencies of the dc magnetisation measured for superconducting $RuSr_2GdCu_2O_8$ led to conclude probable existence of the spontaneous vortex phase (SVP) at temperature proximity of the magnetically determined $T_c^{on}$ [19]. Note, that SVP naturally supported within the layered model, may, however, also be induced for the embedded magnetic inclusions in superconducting volume [20].

The data presented in Fig. 1 reveal considerable differences between the magnetic field dependencies of the resistively measured $T_c^{on}$ for the samples A and B. The data for $Ru_{0.98}Sr_2GdCu_2O_8$ is much less influenced by the magnetic field, and its slope up to the intermediate field values becomes comparable to that characteristic for the samples of $Ru_{0.5}Sr_2GdCu_{2.5}O_{8-d}$ - compare corresponding dependencies established for the samples D and F, which are presented in Fig. 2. The feature may be explained linking the superconducting response of $Ru_{0.5}Sr_2GdCu_{2.5}O_{8-d}$ with that of the superconducting phase induced in $RuSr_2GdCu_2O_8$ and more developed in the $Ru_{0.98}Sr_2GdCu_2O_8$ sample. Within the inhomogeneous picture, for the sample with slightly diminished average Ru/Cu ratio, the regions in which superconductivity nucleates at same reduced temperature $t=T_c/T$ may become larger in size and grow faster out of the scale of the coherence length on lowering the temperature. The magnetic field dependencies of the $T^{90\%Rn}$ are similar for the samples A and B (Fig. 1) but markedly different from those for the two samples of $Ru_{0.5}Sr_2GdCu_{2.5}O_{8-d}$ ( D and F, Fig. 2). This is despite the fact that the $T_c^{on}$ for the sample F was lowered by diminishing its oxygen content to the value comparable to that characteristic of the sample B. It shows that the critical currents for both $Ru_{0.98}Sr_2GdCu_2O_8$ and $RuSr_2GdCu_2O_8$ should be substantially lowered, most probably for presence of the Ru lattice magnetic order, the effect which may also associate larger effective anisotropy of superconducting phase in the parent compound. It would be interesting to investigate the materials with progressively more diluted magnetism in the Ru sublattice for distinguishing if the magnetic order and an effective anisotropy of the superconducting phase remain inherently related in the ruthenocuprate. For best, such investigation should involve suitably substituted single crystals. The growth of even the parent $RuSr_2GdCu_2O_8$ in the crystal form proved, however, quite difficult at least for so far exercised growth conditions.

The difference of approximately 10 K between the $T_c^{on}$ (0) and $T_c^{on\,fl}$ (0) for the samples A and B presents as quite large for the temperature extent of superconducting fluctuations, one should note, however, that such may become enhanced in case of constrained dimensionality of the superconducting phase in $RuSr_2GdCu_2O_8$ (for $YBa_2Cu_3O_7$ , which should have considerably smaller anisotropy coefficient, for its $\xi \approx 10$ Å and $T_c \approx 90$ K the range of fluctuations was estimated to approximately 5 K). For the samples A and B, $T_c^{on\ fl}$ in the high field range remains practically independent of the magnetic field. Figure 3 shows the temperature dependencies of the magnetoresistivity coefficient for $Ru_{0.98}Sr_2GdCu_2O_8$ , which at temperatures above the onset for superconductivity are anomalously lowered with the maximum of the effect at approximately 10 kOe. In view of different superconducting characteristics reported in the literature for different samples of $RuSr_2GdCu_2O_8$ with the same nominal stoichiometry, Fig. 4 compares the temperature dependencies of the resistivity measured for two independently prepared samples B and C of $Ru_{0.98}Sr_2GdCu_2O_8$. The dependencies differ for the slope in the normal state, however the transition onset temperatures (and the $T_c^{R=0}$, here not shown) remain almost unchanged, which confirms that the $T_c$ may be reproducibly stabilized and its value may be considered as characteristic for introducing the diminished Ru/Cu ratio in the synthesis at ambient pressure. These dependencies also show an anomalous lowering of the resistivity with the maximum effect at 10 kOe in both samples – note in the figure the resistivity at 10 kOe, which is lower from that at the zero field in range between 56 K and 51 K for the sample B, and in between 55 K and 48 K for the sample C. A similar anomaly was recently reported for the superconducting sample of $Ru_{0.8}Sn_{0.2}GdCu_2O_8$ [21]. The temperature dependencies presented in Fig. 3 show that the effect onsets at 62 K, which corresponds to the maximum value of $T_c^{on\ fl}$. One may conclude it may originate in phase fluctuations, or in complex combined effect of the magnetic field on the magnetic phase and incipient superconductivity. Note , that the high field diamagnetic fluctuations were also observed in the HTSC superconductors for a comparable range of the magnetic fields above 10 kOe [22]. The magnetic field dependencies of the $T_c^{\,on}(H)$, $T_c^{on\ fl}(H)$ for $Ru_{0.5}Sr_2GdCu_{2.5}O_{8-d}$ (Fig. 2, sample F) also reveal modest increase of the $T_c^{on\ fl}$ and $T_c^{on}$ at the magnetic fields smaller than 10 kOe. Unfortunately, since we failed to observe in this sample the superconducting anomaly in specific heat (in similar to outcome of the specific heat measurements for $Ru_{0.98}Sr_2GdCu_2O_8$ sample B, see [11]), the effect could not be approached in the bulk thermodynamic measurement for which influence of the superconducting filaments, as opposed to the main superconducting phase, could be excluded.

Before commenting the features of normal resistive state in the samples A and B, it is worth noting that the multi-component magnetism of $RuSr_2GdCu_2O_8$ originates in the magnetic sublattices of both the Gd and Ru ions (for the ac susceptibility-temperature dependencies for the samples A and B see Ref. 11, there Fig. 4.b). The commonly reported *M(H)* dependences are hysteretic at temperatures below the magnetic transition temperature $T_m^{Ru}$ and resemble the magnetic response of a weak ferromagnet. The overall magnetisation of the compound is however build of the dominating response of Gd moments (the Gd sublattice orders antiferromagnetically at $T_N^{Gd}$=2.5 K [23]) and only adds the component due to magnetism of the ordered sublattice of Ru, two sublattices being most probably polarised with their resulting effective internal fields [24]. The soft hysteresis present in the *M(H)* dependences below $T_m^{Ru}$ should then originate in an effective ferromagnetic component on antiferromagnetically aligned sublattice of the Ru moments [23,25].

Temperature dependencies of the resistivity of $RuSr_2GdCu_2O_8$ and $Ru_{0.98}Sr_2GdCu_2O_8$ also reveal the features which associate with the Ru magnetic ordering – there is an insignificant decrease of the resistivity present at temperatures close to $T_m^{Ru}$. By the temperature correspondence the feature seems reflecting an increase of conductivity in the structural layers of the Ru sublattice upon entering its magnetically ordered state. The relative extent of the decrease of resistivity at $T_m^{Ru}$ remains a bit sample dependent, however it always associates with the temperature onset for the negative magnetoresistivity observed in broad range of temperatures below $T_m^{Ru}$ [18]. The noted feature resembles that of the resistivity-temperature dependence of $SrRuO_3$ itinerant ferromagnet at temperatures just below its ferromagnetic transition at $T_C$≈160-163 K. In Fig. 5 we present the magnetic field dependencies of these characteristic temperatures established for the samples A ($RuSr_2GdCu_2O_8$) and B ($Ru_{0.98}Sr_2GdCu_2O_8$). The temperatures were determined as the upper temperature limit for fitting the derivative of resistivity with the linear temperature dependence. The triangles in this figure show in matching dependence the characteristic temperatures obtained for the sample A from the specific heat data. At these temperatures the $C_p(T)/T$ departs downward from its linear increase on lowering the temperature, which is characteristic for quite broad range above $T_m^{Ru}$ (the corresponding temperature dependencies of the specific heat for samples A and B were presented in Ref.18, there Fig. 4). The resistive anomaly for sample B ($Ru_{0.98}Sr_2GdCu_2O_8$) occurs at slightly lower temperatures than for sample A (the result holds despite rather large uncertainty in determination of the temperatures - see dotted lines in Fig. 5). As for comparing the samples with different ratio of Ru/Cu the effect would support presence of the magnetic dilution in the magnetically responding lattice of the Ru moments. The positive field shift of such resistively determined temperatures suggests about the ferromagnetic character of probed magnetic interactions. The shift of zero field values between samples A and B also complies with the $C_p(T)/T$ data (Ref. 18, there Fig. 4). Similar behaviour of the magnetoresistivity of strontium ruthenate may suggest similar character for the involved Ru magnetism for an effective metallic conduction channel present in the investigated samples. Note that when searched for the origin of the spatial phase separation in the Ru1222-type ruthenocuprate, the measurements of magnetisation in the normal state hinted that the magnetic response may involve that of well dispersed magnetic islands of $SrRu_{1-x}Cu_xO_3$, then be present in the investigated material [2]. Here reported features, however, may originate in the Ru magnetism of the 1212-type phase.

Figure 6 presents the temperature dependence of the dc magnetisation measured for sample D of $Ru_{0.5}Sr_2GdCu_{2.5}O_{8-d}$ at the magnetic field of 500 Oe (open circles). The substantial re-entrant like increase of the magnetisation observed at low temperatures in the superconducting state may be interpreted as reflecting the unscreened contribution of paramagnetism of the $Gd^{3+}$ ions. The inset in Fig. 6 shows the dc magnetisation measured at 42 Oe and at 70 kOe (open circles and the line, respectively). There shown the FCW (field cooled on warming) branch of magnetisation at the low field of 42 Oe remains flat down to almost 2 K as it would be expected for the magnetic response typical of Meissner state. At 70 kOe the magnetization remains paramagnetic in a whole range of the accessed temperatures (line in the inset to Fig. 6). The solid squares in the inset, which match that dependence, compare the dc susceptibility for the non-superconducting $GdCu_2Ba_3O_{6.2}$. This directly provides that the magnetisation of $Ru_{0.5}Sr_2GdCu_{2.5}O_{8-d}$ in the high field limit matches the expected magnetic response of the paramagnetic $Gd^{3+}$ sublattice in the sample. In [13] it was noted that the overall magnitudes of high field magnetisation (at 70 kOe, 4.5 K) probed for several samples of $Ru_{1-x}Sr_2GdCu_{2+x}O_{8-d}$ do not scale with shift in x. Figure 7 presents the hysteretic *M(H)* dependence measured for sample D at 4.5 K i.e. in its superconducting state. Since the hysteretic behaviour ceases to exist at $T_c$ (not shown) it should reflect the superconducting irreversibility. The lower inset in this figure shows the magnified low field range for presenting the virgin part of the *M(H)* dependence, which features the local minima. The temperature dependence of the magnetic field values for lower minimum, taken for correspondence to the first penetration field of superconducting phase are presented in the upper inset to this figure.

The qualitatively similar *M(H)* dependencies were reported for the series of $Ru_{1-x}Sr_2GdCu_{2+x}O_{8-d}$ for samples with several different values of the parameter *x* [13, 26]. It is interesting to note that whereas the superconducting $T_c$ attains maximum for approximately x=0.5, the low temperature magnetisation show the contribution of the Gd paramagnetism which is progressively diminished with increase of the x parameter [26] (commented are the samples prepared at the same conditions – note that whereas the effective charge doping should depend both on the Ru and oxygen concentration, the activation energy for modifying the latter, presumably in the Ru/Cu –O structural chains, may depend on the parameter x). It is worth to comment that the $YBa_2Cu_3O_{7-d}$ (YBCO) superconductor, which has similar 123-type crystal structure, became interesting in the HTSC family for its relatively low effective anisotropy - comparing to $Bi_2Sr_2CaCu_2O_{8+d}$ for which the spacing of Cu-O planes is larger only by 3 Å (12 Å vs. 15 Å), the anisotropy parameter $\gamma^2=m_c/m_a$ for YBCO is smaller at least by the order (5-7 vs. 50-200). Fundamental difference provides presence of the Cu-O chains in the 123-type structure, which are situated in the so called charge reservoir layer (same distant to the two $CuO_2$ planes) and which Cu-O bonds are believed to support the superconducting condensate thus diminishing the superconducting anisotropy. For the series of $Ru_{1-x}Sr_2GdCu_{2+x}O_{8-d}$ by changing the level of Cu->Ru substitution the compounds seem progressively evolving from the limit of $RuSr_2GdCu_2O_8$ M/SC layered hybrid (note the *I-V* characteristics for the single crystals [6]) toward much less electronically anisotropic 123-type like structure. It may be of that process of gaining the isotropy of superconducting phase to be reflected in the gain of the screening of the paramagnetic response of the Gd layers, which, however, are structurally closest to and positioned in between two neighbouring Cu-O planes. If the c axis anisotropy of the superconducting order parameter should remain universally sensitive to the layering distances in the cuprate structures, it would be conceptually tempting to investigate for the structural association of incipient superconductivity in such structures.

Figure 8 presents the *M(H)* dependence which was calculated by subtracting the paramagnetic dependence measured for the non-superconducting $GdBa_2Cu_3O_{6.2}$ from the *M(H)* data presented in Fig. 7. Then, the figure should approximate the superconductivity caused component in M(H) of $Ru_{0.5}Sr_2GdCu_{2.5}O_{8-d}$ and the local minimum at low fields may reflect the field at which first vortices move into the superconducting volume. This minimum and the second broad minimum at the higher field values may then reflect the effective spatial anisotropy of the compound's superconducting phase (the experiment provides an averaged moment for the randomly aligned crystallites in the magnetic field). The dependence in Fig. 8 may, however, also reflect a multi-component nature of the measured signal for possibly larger scale spatial inhomogeneity of the superconducting response. This sample was investigated by means of the electron microscopy and Fig. 9 presents the topological pictures and the compositional scan of its selected area. Since the scans for all involved elements showed the compositional inhomogeneity at the micrometer range one should be aware of the influence of the inclusions which carry different level of cationic substitutions, or the presence of other related impurity phases, which high level of dispersion could effectively preclude reliable estimation of the volume in a diffraction experiment. In principle, an effective inhomogeneity could also be developed by stacking together blocks of the structural layers which associate similar Ru-Cu substitution on the scale insensitive to diffraction, however relevant for supporting superconducting condensate. This would resemble the postulate which was based in the local scale structural studies of the superconducting sample of $RuSr_2GdCu_2O_8$ [5]. Further investigation of differently scaled structural features seem be required also for $Ru_{0.5}Sr_2GdCu_{2.5}O_{8-d}$ .

Since in the Cu substituted ruthenocuprate the oxygen content becomes defining the superconducting $T_c$, its differences and uniformity on the scale defined by the Ru/Cu substitution should be considered. Figure 6 presents the paramagnetic dependence of the dc magnetisation measured at 500 Oe for the non-superconducting sample E, which was formed of the sample D after depleting some oxygen by annealing at the reducing conditions. Part of this material was re-oxygenated, which re-established the superconducting phase – the sample F (details were provided in the chapter on synthesis). Figure 2 compares the characteristic temperatures of the resistive superconducting transitions measured for the samples D and F. Besides the expected lowering of superconducting $T_c$ [14], the temperature interval between $T_c^{on}$ and $T_c^{on\ fl}$ gets considerably smaller for the re-oxygenated sample, what might reflect better homogeneity of the superconducting phase in the material.

**Conclusions**

The report comments on the characteristics of the superconducting transitions in $RuSr_2GdCu_2O_8$, $Ru_{0.98}Sr_2GdCu_2O_8$ and $Ru_{0.5}Sr_2GdCu_{2.5}O_{8-d}$ ruthenocuprates. The differences in the magnetic field dependencies of the superconducting onset temperatures ($T_c^{on}$) are interpreted primarily for an inhomogeneous nucleation of the superconducting phase in the parent ruthenocuprate. The $dH_{c2}(T)/dT$ ratios for $RuSr_2GdCu_2O_8$ and $Ru_{0.98}Sr_2GdCu_2O_8$ are found considerably lower than for the two samples of $Ru_{0.5}Sr_2GdCu_{2.5}O_{8-d}$, even one of which has comparably low $T_c^{on}$ by diminishing its oxygen content. The data seems to support the modification of the effective anisotropy of the superconducting phase in the ruthenocuprate with proceeding of Ru/Cu substitution. The field induced anomalous lowering of the resistivity just above the superconducting onset temperature $T_c^{on}$ , investigated in $Ru_{0.98}Sr_2GdCu_2O_8$ , may involve enhanced superconducting fluctuations. Despite that the anisotropy of the superconducting phase in $Ru_{1-x}Sr_2GdCu_{2+x}O_{8-d}$ seems to gradually diminish with progression of the Cu-Ru substitution, the superconducting characteristics of $Ru_{0.5}Sr_2GdCu_{2.5}O_{8-d}$ ($T_c$=72 K) yet reveal a significant contribution of the Gd paramagnetic signal at low temperatures. The properties of the Ru-Cu substituted phases of the ruthenocuprates may thus present an opportunity to probe the cuprate superconductivity in the compound structurally similar to YBCO, which supports significantly enhanced superconducting anisotropy and comparatively high $T_c$.

**Acknowledgements**

PWK thanks Dr. Russ Cook of the Materials Science Division of Argonne National Laboratory for providing with the electron microscopy analysis of the samples investigated within the NSF Grant DMR-0105398, the SEM pictures of which are included in Fig. 9. Part of the research was financed by the Polish Ministry of Science and Higher Education research project funding for the years 2007–2010.

**Figures**

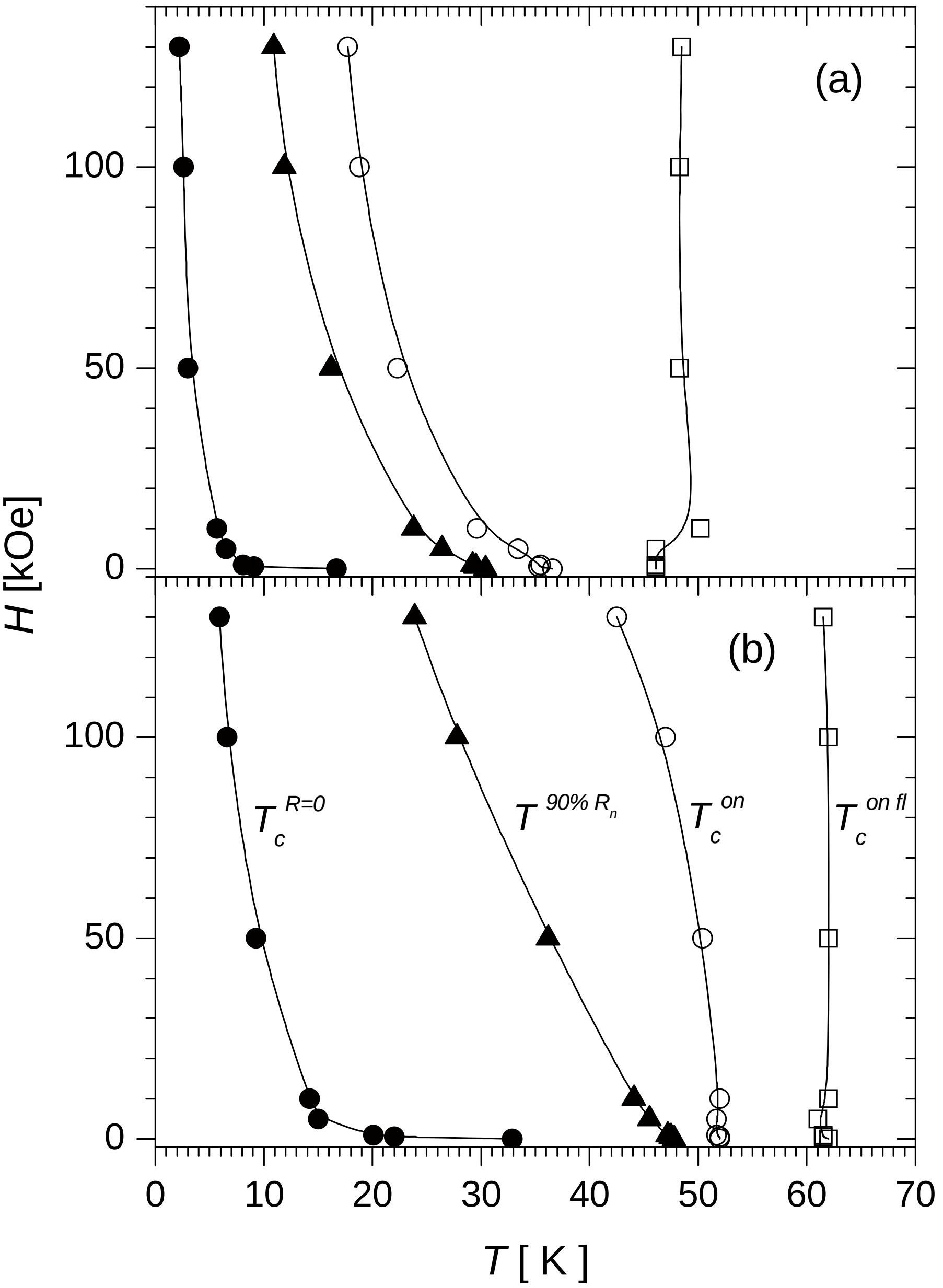

**Figure 1.** The magnetic field dependencies of the temperatures characteristic for the resistive superconducting transitions in (a): $RuSr_2GdCu_2O_8$ (sample A) and (b): $Ru_{0.98}Sr_2GdCu_2O_8$ (sample B). $T_c^{R=0}$ – temperature of the zero resistivity state, $T^{90\% Rn}$ – temperature corresponding 10% decrease of the resistivity relative to its value at $T_c^{on}$, $T_c^{on}$ – temperature of the onset of superconducting transition, $T_c^{on\,fl}$ – limiting temperature associated with incipient superconductivity, see text for discussion.

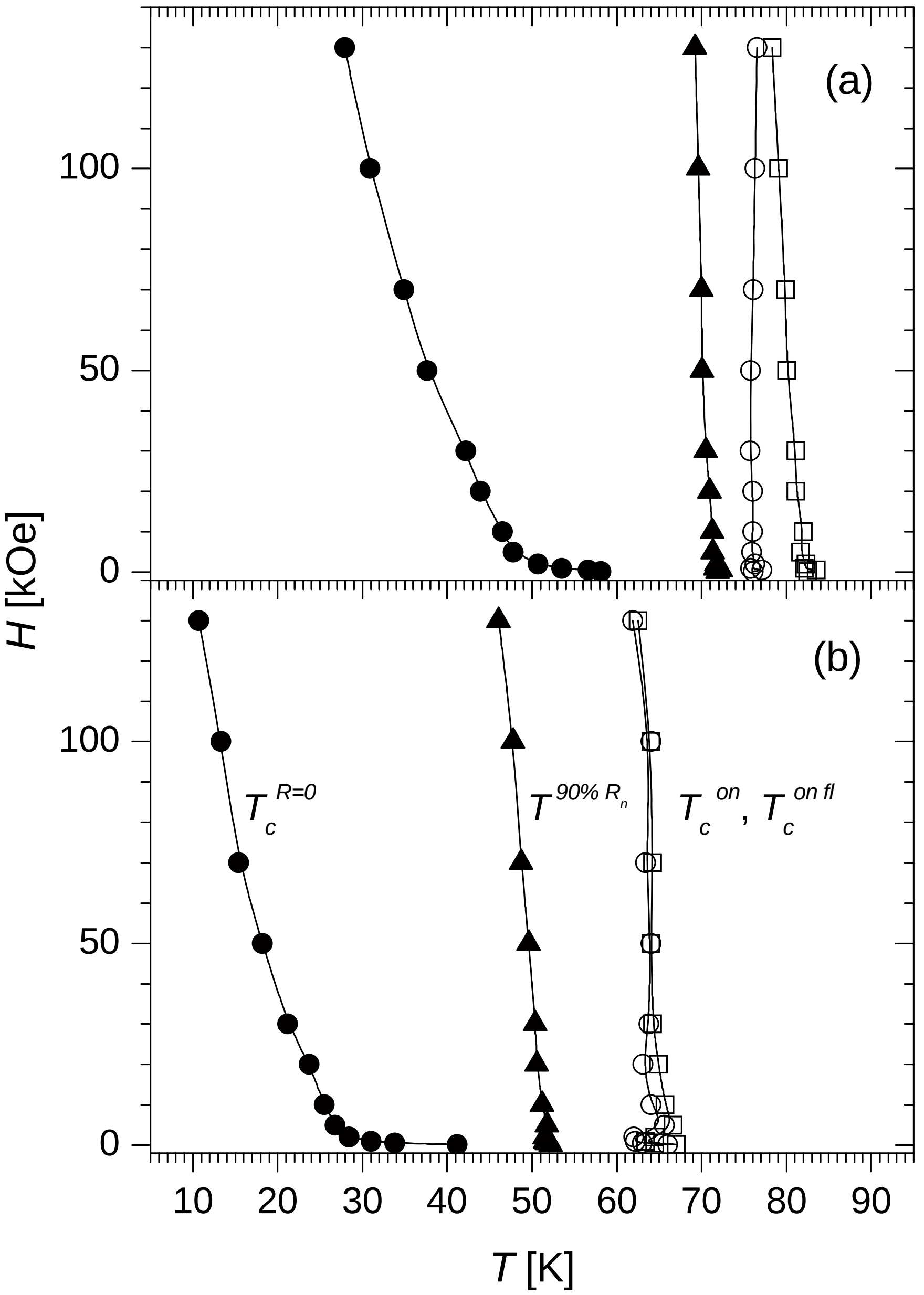

**Figure 2.** The magnetic field dependencies of the temperatures characteristic for the resistive superconducting transitions in $Ru_{0.5}Sr_2GdCu_{2.5}O_{8-d}$ (a) sample D and (b) sample F. $T_c^{R=0}$ – temperature of the zero resistivity state, $T^{90\%\,Rn}$ – temperature corresponding 10% decrease of the resistivity relative to its value at $T_c^{on}$, $T_c^{on}$ – temperature of the onset of superconducting transition, $T_c^{on\,fl}$ – limiting temperature associated with incipient superconductivity, see text for discussion.

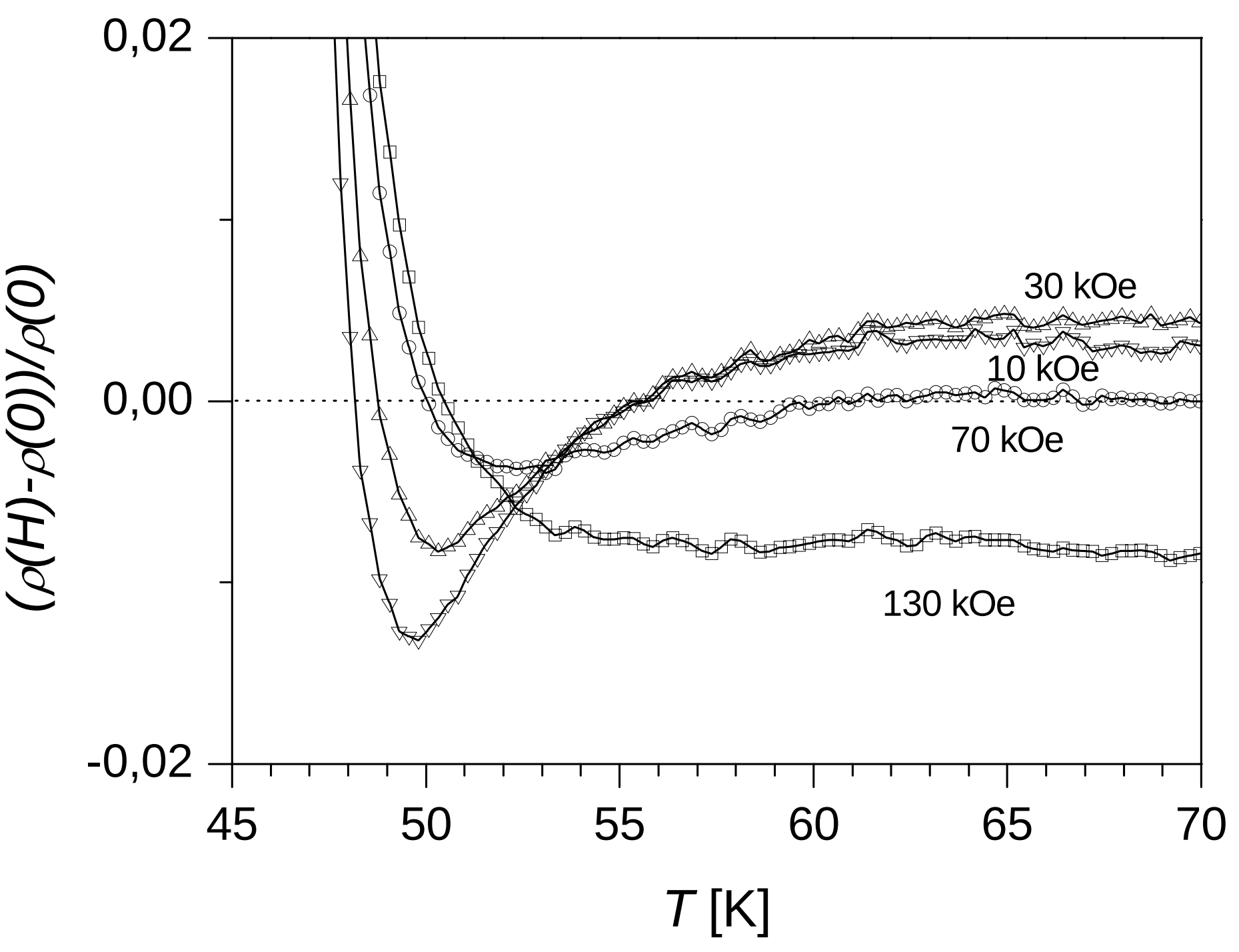

**Figure 3.** The temperature dependences of the magnetoresistivity in a vicinity of $T_c^{on}$ for $Ru_{0.98}Sr_2GdCu_2O_8$ sample C. Negative magnetoresistivity in the normal, and magnetically ordered, state is reflected in lower values (to right of the figure) of the magnetoresistivity coefficient for 70 kOe and 130 kOe as opposed to its values calculated for 10 kOe and 30 kOe.
Note that for the 130 kOe curve the increase at 52 K appears in consequence of $T_c^{on}$(H=0)≈52 K i.e. does not mark the $T_c^{on}$ (H=130 kOe).

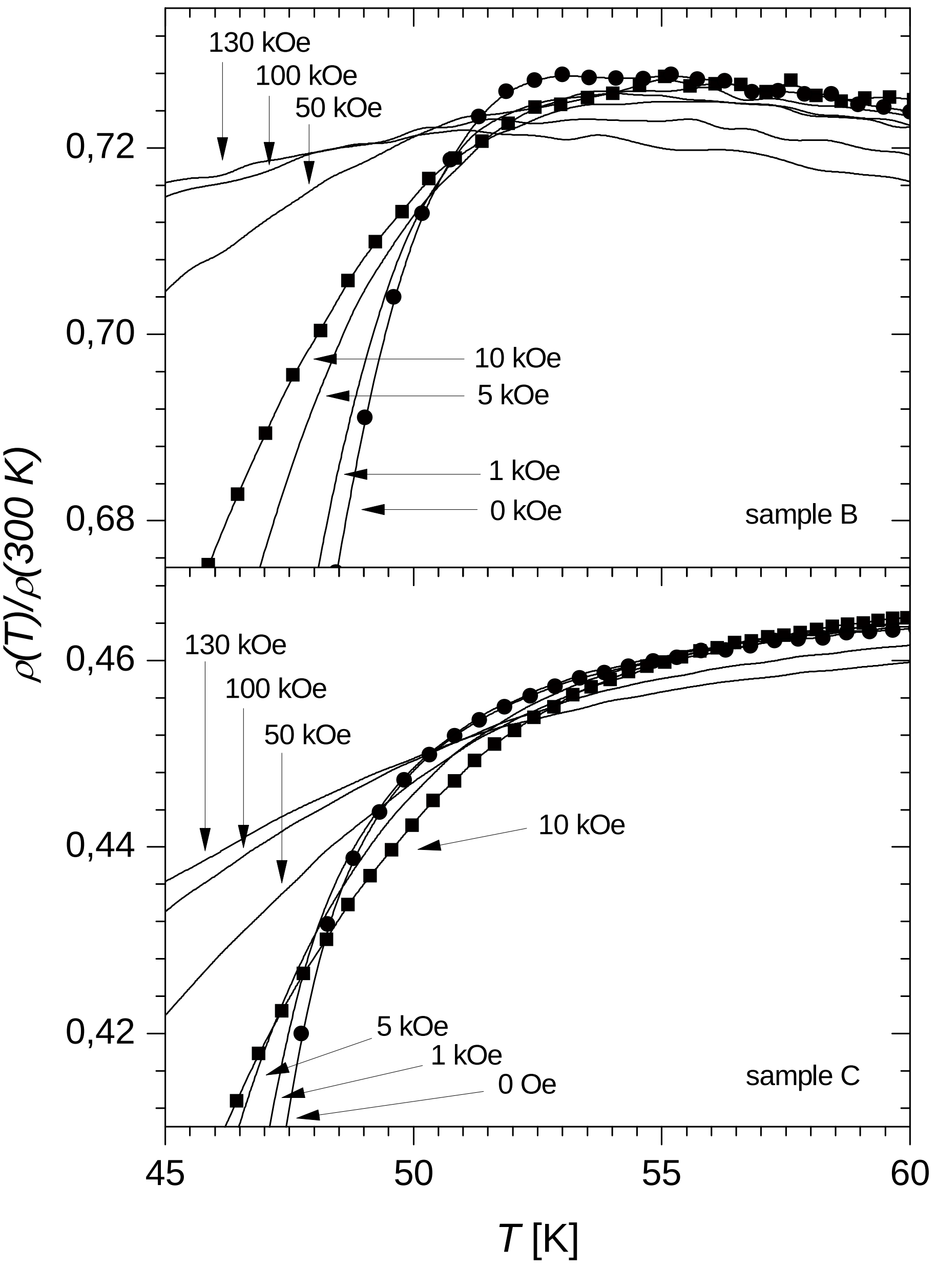

**Figure 4.** The temperature dependencies of the resistivity measured at different values of the magnetic field close to the onset temperature of superconducting transition for $Ru_{0.98}Sr_2GdCu_2O_8$ sample B (upper) and sample C (lower). Circles and squares emphasise the dependencies measured at 0 and 10 kOe, respectively.

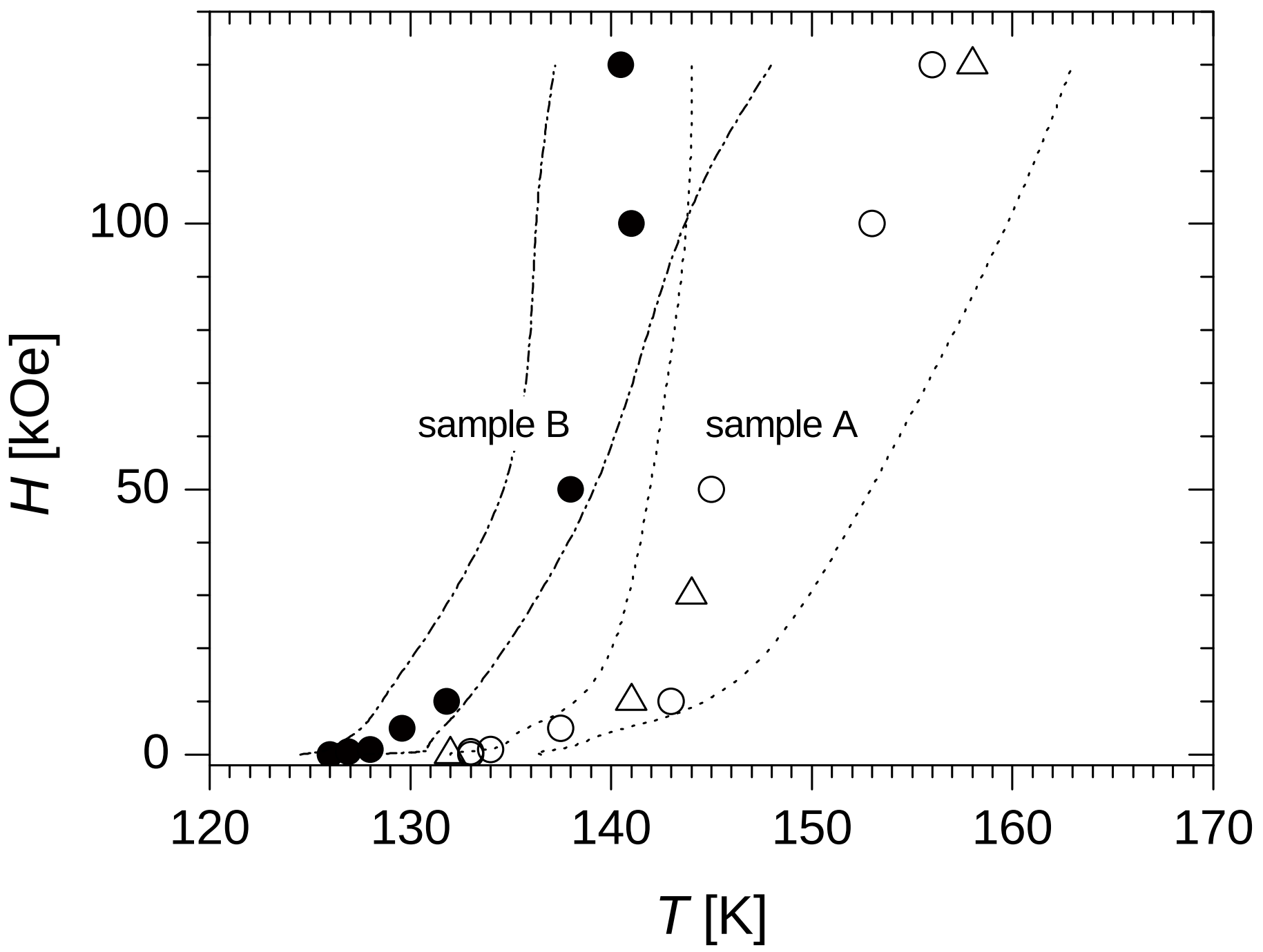

**Figure 5.** The magnetic field dependencies of the temperatures characteristic for the resistivity anomaly in a vicinity of $T_m^{Ru}$ for two simultaneously prepared samples: $RuSr_2GdCu_2O_8$ (sample A) – open circles, and $Ru_{0.98}Sr_2GdCu_2O_8$ (sample B) – closed circles. Dotted lines provide for uncertainty of estimation. The corresponding characteristic temperatures read of the temperature dependencies of the specific heat for sample A are shown with triangles (see text).

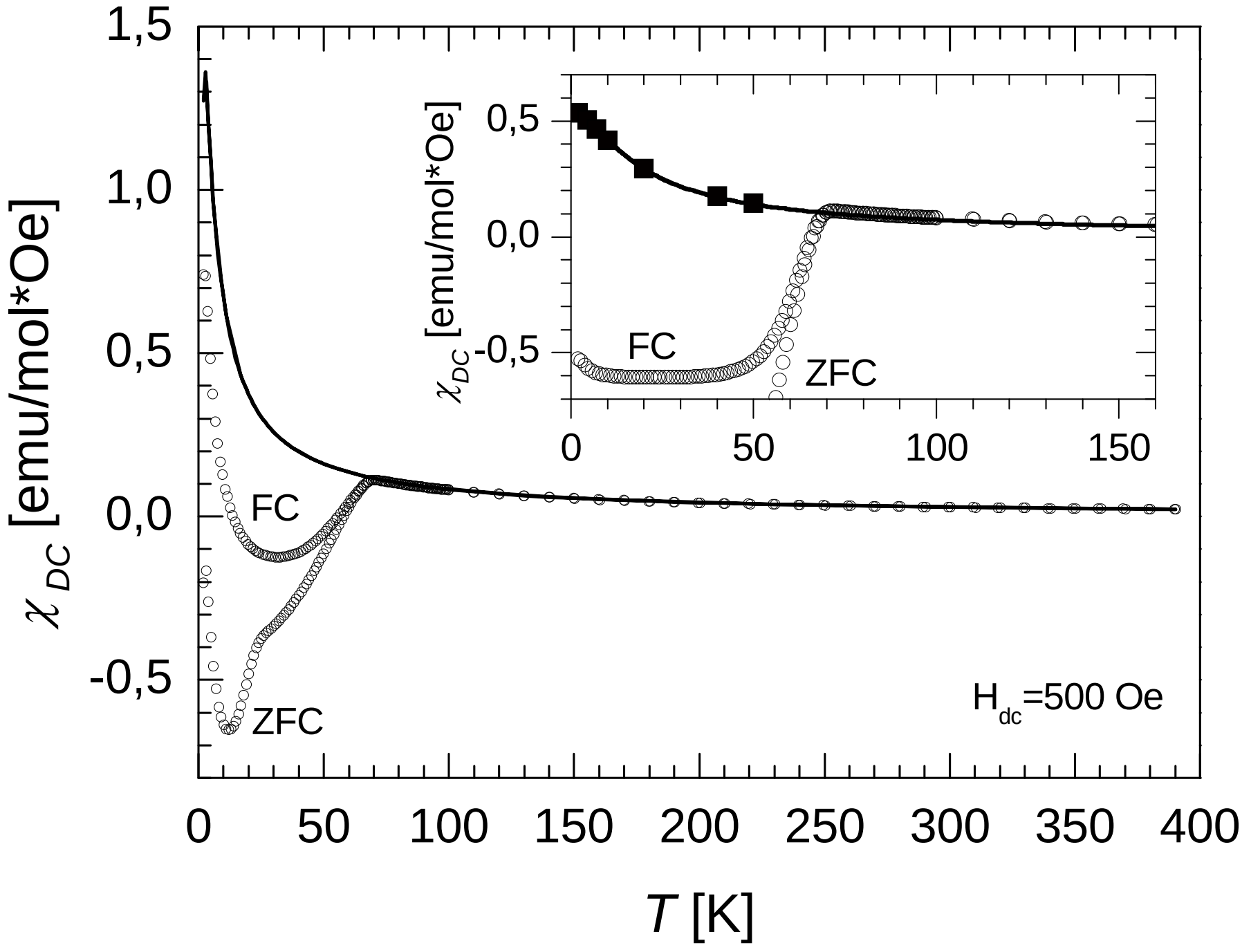

**Figure 6.** The temperature dependencies of the dc susceptibility for $Ru_{0.5}Sr_2GdCu_{2.5}O_{8-d}$ at 500 Oe: open circles - sample D, solid line - sample E. Inset shows the dc susceptibility for sample D measured for two magnetic field values: open circles – at 42 Oe, solid line - at 70 kOe. Solid squares in the inset (which follow the line) compare the values of magnetic susceptibility of $GdBa_2Cu_3O_{6.2}$ measured also at 70 kOe.

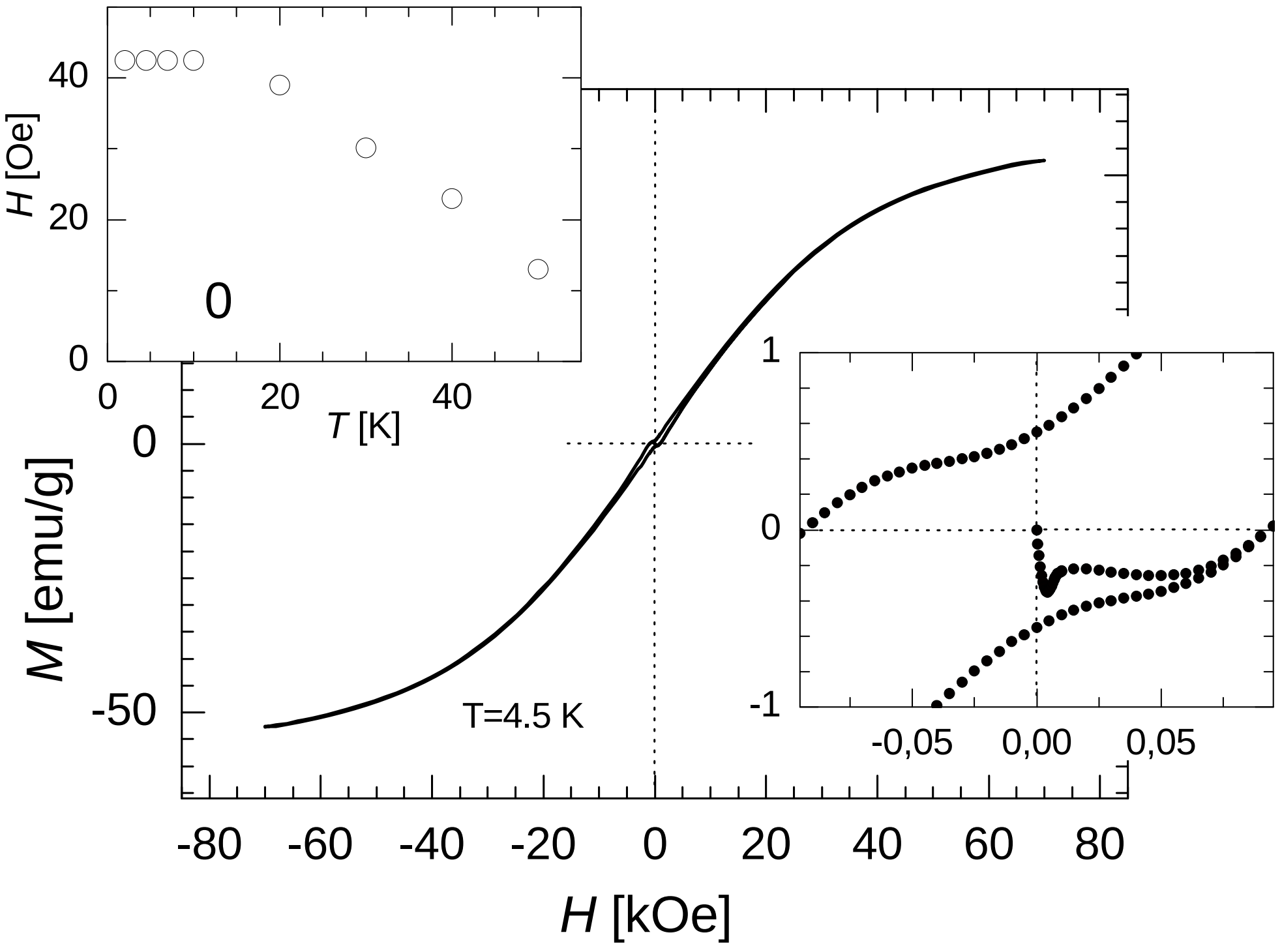

**Figure 7.** The hysteretic magnetic field dependence of the magnetisation measured for $Ru_{0.5}Sr_2GdCu_{2.5}O_{8-d}$ sample D at 4.5 K. Lower inset presents the magnified origin part of the M(H) plot. Upper inset shows the magnetic field values, which correspond to the lower field local minimum in the *M(H)* dependences measured at different temperatures.

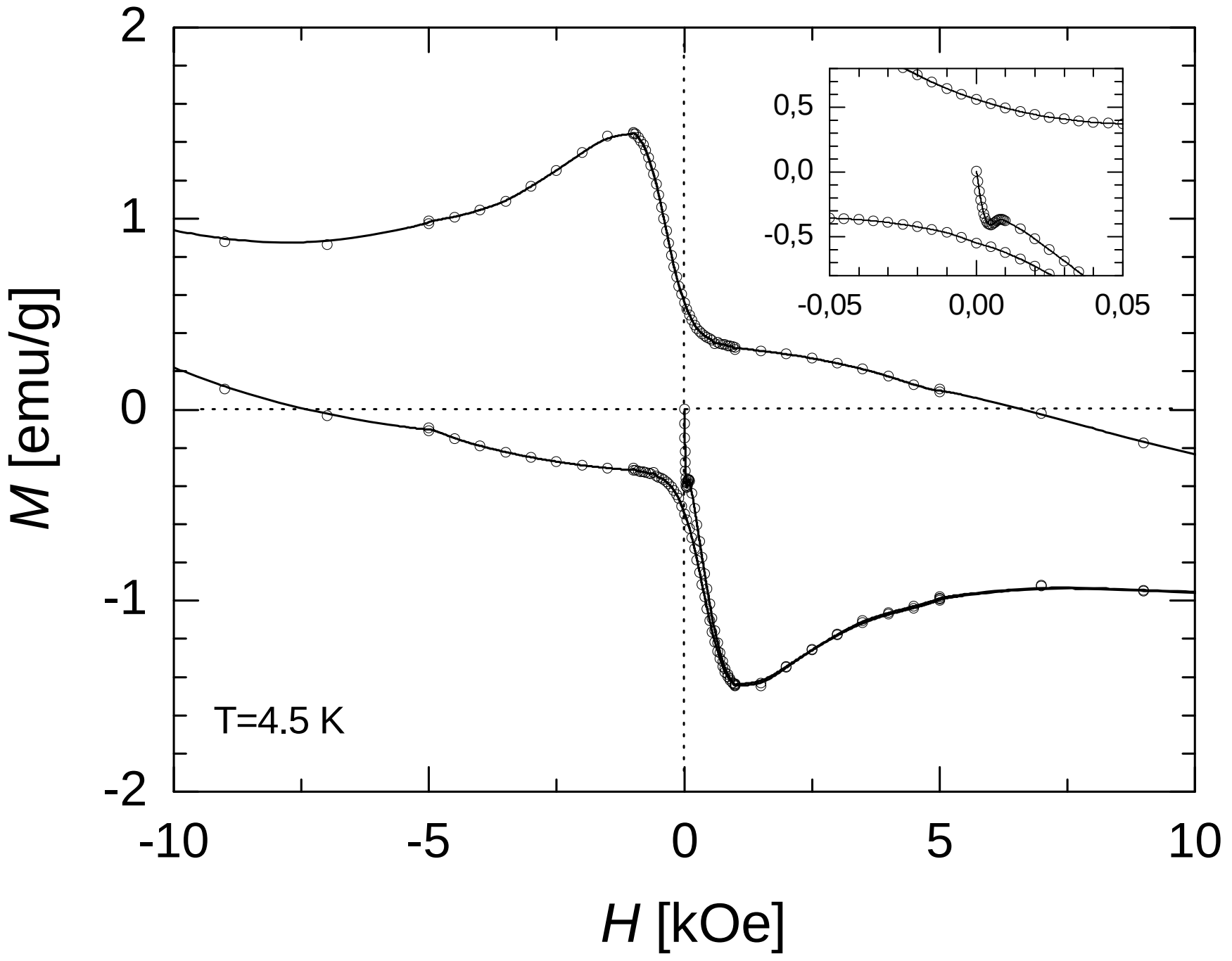

**Figure 8.** The superconductivity caused irreversible contribution to the *M(H)* dependence at 4.5 K calculated for sample D of $Ru_{0.5}Sr_2GdCu_{2.5}O_{8-d}$. Data comes in subtraction of the measured paramagnetic response of $GdBa_2Cu_3O_{6.2}$ from the experimental data presented in Fig. 7.Inset shows the local minimum at low field.

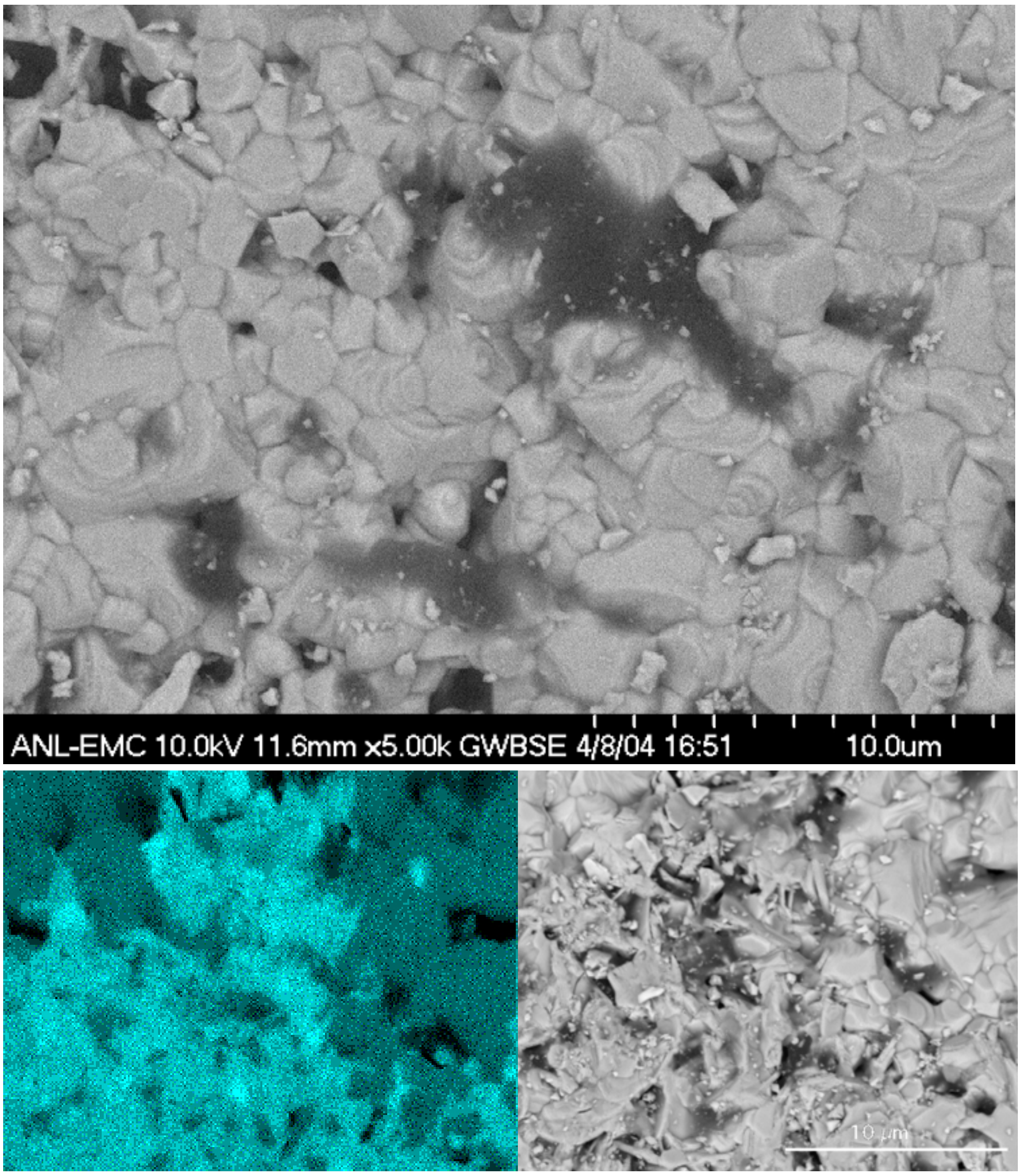

**Figure 9.** Colour online. SEM pictures of two selected areas of the $Ru_{0.5}Sr_2GdCu_{2.5}O_{8-d}$ sample D. The upper and lower right views were acquired using a dedicated backscatter electron detector in the composition mode. Lower left is the EDXS Sr-L X-ray net intensity map for the area presented in the lower right image. The map was formed by scanning the image pixel-by-pixel and collecting the EDXS spectrum at each pixel point. The incident electron interaction volume extends over approximately four pixels. Differences in brightness correspond to not uniform distribution of the element. The maps done for other elements essentially duplicate or reverse here shown light vs. dark areas revealing more Cu and Gd in the upper left and right and more Sr and Ru in the lower left to the middle top. Additionally there were found spots of Al and few more distributed areas with presence of carbon (not shown).